\documentclass[conference,compsoc]{IEEEtran}

\usepackage{cite}

\usepackage[dvipdfmx]{graphicx}
\usepackage{mediabb}

\usepackage{amsmath}

\usepackage{url}
\usepackage[pagebackref=true,breaklinks=true,colorlinks,bookmarks=false]{hyperref}

\usepackage{times}
\usepackage{multirow}

\usepackage{xspace}
\makeatletter
\DeclareRobustCommand\onedot{\futurelet\@let@token\@onedot}
\def\@onedot{\ifx\@let@token.\else.\null\fi\xspace}
\def\etal{\emph{et al}\onedot}

\begin{document}
%
\title{%
  Single-epoch supernova classification with deep convolutional neural networks
}



%
\author{
  \IEEEauthorblockN{%
    Akisato Kimura\IEEEauthorrefmark{1},
    Ichiro Takahashi\IEEEauthorrefmark{2},
    Masaomi Tanaka\IEEEauthorrefmark{3},
    Naoki Yasuda\IEEEauthorrefmark{2},
    Naonori Ueda\IEEEauthorrefmark{1} and
    Naoki Yoshida\IEEEauthorrefmark{2}
  }
  \IEEEauthorblockA{\IEEEauthorrefmark{1}
    NTT Communication Science Laboratories,
    Keihanna Science City, Kyoto, 619--0237 Japan.
    Email: akisato@ieee.org
  }
  \IEEEauthorblockA{\IEEEauthorrefmark{2}
    Kavli IPMU, the University of Tokyo,
    Kashiwa, Chiba, Japan.
    Email: ichiro.takahashi@ipmu.jp
  }
  \IEEEauthorblockA{\IEEEauthorrefmark{3}
    National Astronomical Observatory of Japan,
    Mitaka, Tokyo, Japan.
    Email: masaomi.tanaka@nao.ac.jp
  }
}


\maketitle

\begin{abstract}
  Supernovae Type-Ia (SNeIa) play a significant role in exploring the history of the expansion of the Universe, since they are the best-known standard candles with which we can accurately measure the distance to the objects.
  Finding large samples of SNeIa and investigating their detailed characteristics have become an important issue in cosmology and astronomy.
  Existing methods relied on a photometric approach that first measures the luminance of supernova candidates precisely and then fits the results to a parametric function of temporal changes in luminance.
  However, it inevitably requires multi-epoch observations and complex luminance measurements.
  In this work, we present a novel method for classifying SNeIa simply from single-epoch observation images without any complex measurements, by effectively integrating the state-of-the-art computer vision methodology into the standard photometric approach.
  Experimental results show the effectiveness of the proposed method and reveal classification performance comparable to existing photometric methods with multi-epoch observations.
\end{abstract}

%
\IEEEpeerreviewmaketitle


\section{Introduction}
\label{sec:intro}

  In physical cosmology and astronomy, 
  dark energy is the best-accepted hypothesis explaining observations obtained since 1990s indicating that the Universe is expanding at an accelerating rate.
  High-precision measurements of the expansion of the Universe are required to understand how the expansion rate changes over time.
  In general relativity, the evolution of the expansion rate is parameterized by the cosmological equation of state (the relationship between temperature, pressure, and combined matter, energy, and vacuum energy density for any region of space).
  Measuring the equation of state for dark energy is one of the main challenges currently facing observational cosmology.


 \begin{figure}[t]
  \begin{center}
    \includegraphics[mediaboxonly,width=0.985\linewidth]{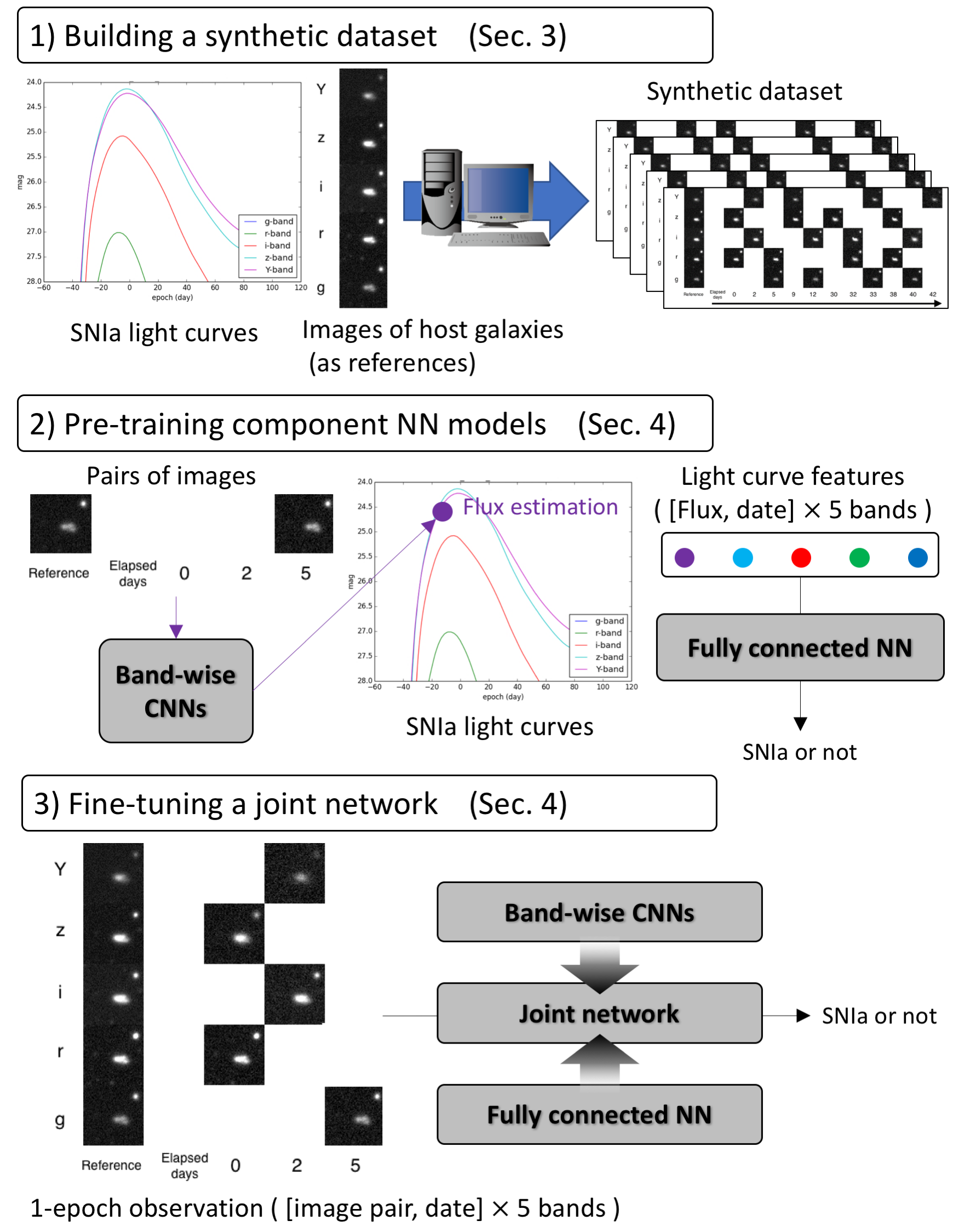}
  \end{center}
  \caption{Our proposed approach}
  \label{fig:approach}
\end{figure}

  Supernovae are useful for this purpose because they are excellent standard candles that are objects for which the intrinsic brightness is known.
  In particular, Type Ia (``one-A'') supernovae (called SNeIa below) are the best-known standard candles across cosmological distances because of their extreme and consistent luminosity \cite{Riess1998,Perlmutter1999}.
  By looking at the relationship between the distance to an object and its redshift (the phenomenon whereby the wavelength of the light from an object is shifted toward longer wavelength) that can be measured by spectroscopy, we can measure the history of the expansion of the Universe.
  Finding increasingly large samples of SNeIa and investigating the detailed characteristics of these supernovae have become an important research task in physical cosmology and astronomy.

\begin{figure}[t]
  \begin{center}
    \includegraphics[mediaboxonly,width=0.985\linewidth]{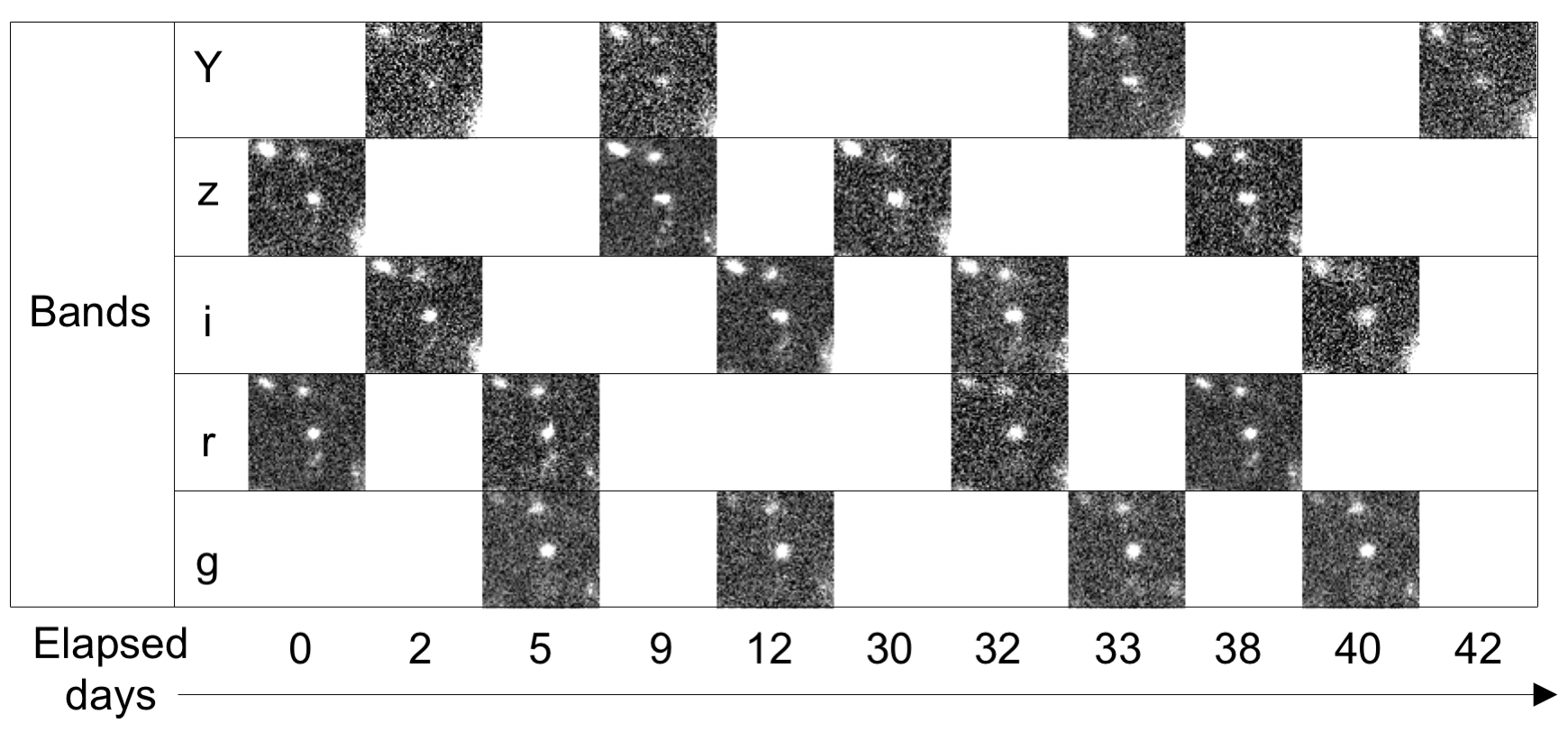}
    \end{center}
  \caption{Example of observations}
  \label{fig:observation}
\end{figure}

  One of the current standard pipelines of supernova detection is as follows:
  (1) First, a sky image is obtained using a telescope with broad-band filters, which are denoted for example g, r, i, z and y filters dependign on the wavelengths.
  The broad-band filter combinations are scheduled in advance so that 1 to 3 band images are taken on every observation date and all the bands have a similar number of images, and thus we obtain a predefined number of images for every band on the predefined schedule if the weather permitted.
  Figure \ref{fig:observation} is an example of observations focused on a certain astronomical object. 
  (2) Then, transient object candidates are detected by subtracting the obtained image from a reference image convoluted with an appropriately optimized filter to match the image quality. 
  (3) All the detected candidates are checked \emph{manually} to determine whether or not each candidate is an SNIa.
  (4) Selected SNIa candidates are employed in spectroscopic follow-up observations to check whether or not the candidate is actually an SNIa and to investigate their parameters.

  As described above, the process of detecting supernovae relies heavily on human effort, however, at most only 100 of over $10^7$ candidates can proceed to follow-up spectroscopic observations.
  Automatic and precise identification of SNIa samples is recognized as an urgent need for cosmologists and astronomers.
  Most of the existing methods adopted a photometric approach (e.g. \cite{Sullivan2006}) that first measures the luminance (called flux in cosmology) of individual supernova candidates and then fits the results to a parametric representation of light curves (temporal changes in flux).
  However, most of the approaches require precise and complex flux measurements and observations at many epochs
  \footnote{In astronomy, the term ``epoch'' or formally ``observation epoch'' refers to a moment in time used as a reference point for some time-varying astronomical quantity.}
  for fitting them to the light curve templates.
  For example, a famous dataset released by the Supernova Photometric Classification Challenge (SNPCC) \cite{Kessler2010a} contains 4 to 40 flux measurements in each band.  Many existing methods employ this dataset as a benchmark, and thus they inevitably utilize at least 4 observations in each band.
  However, the duration for which the luminance of SNeIa has been sufficiently bright to capture them with a telescope is at most 2 months, and we have to complete both photometric broad-band and spectroscopic follow-up observations within this period.
  This means that broad-band surveys should be completed as soon as possible for rapid follow-up observations.

  This paper proposes a novel method for selecting SNeIa from a given set of supernova candidates with only a single-epoch observation image for each band.
  Our method shown in Figure \ref{fig:approach} effectively employs the state-of-the-art computer vision methodology, and integrates it into the standard photometric approach.
  Specifically, our method first builds a convolutional neural network (CNN) to estimate the flux of a supernova candidate from a pair of telescope images, and then constructs another fully connected neural network (NN) for the classification, where the estimated fluxes and observation dates of 5 bands are used as features for classification.
  The models for flux estimation and classification are both based on neural networks, which means that the joint network that classifies supernovae directly from images can be fine-tuned from separately pre-trained networks.
  To enable the separate pre-training of component networks, we synthetically build a new large-scale dataset consisting of a type of supernova, simulated multi-epoch observations and corresponding fluxes, described in Section \ref{sec:data}.

  The efficient method proposed in this paper will be useful for a larger US-led survey by the Large Synoptic Survey Telescope (LSST) in the near future, which is expected to discover more than 200K SNeIa every year \cite{LSSTScienceCollaboration2009}.




\section{Related work}
\label{sec:related}

  Supernova classification has two main steps:
  (1) selecting only transient objects from a given set of candidates, and
  (2) identifying a type of supernova from the remaining candidates.
  This paper aims at the second step. 

  The process of detecting supernovae relies heavily on human effort.
  The problem here is that only 0.1\% of vast amount of candidates are actual transient objects and the others are ``bogus'' (fake transient objects).
  This is because
  (1) filter optimization for subtracting the obtained image from a reference image often fails, yielding fake transients.
  (2) cosmic ray hits also yields fake detections.
  Based on this background, some works have started to introduce image processing and machine learning techniques.
  Bailey \etal \cite{Bailey2007}, Bloom \etal \cite{Bloom2012} and Brink \etal \cite{Brink2013} applied random forests and achieved a true positive rate (TPR) of 92.3\% at a false positive rate (FPR) of 1.0\%.
  Morii \etal \cite{Morii2016} first introduced deep neural networks for this purpose and achieved FPR of 0.85\% at TPR of 90.0\%.

  Identifying SNeIa samples from the remaining candidates are also significant in astronomy, since at most only 100 of the candidates can proceed to follow-up spectroscopic observations.
  In contrast to bogus rejection, very little of the previous research on identifying a type of supernovae employed image recognition techniques.
  Instead, they relied on a photometric approach that first measures the flux of supernova candidates for several filters and then fits them to a parametric representation of light curves.
  Sullivan \etal \cite{Sullivan2006} may have been the first to use this approach. 
  Kessler \etal \cite{Kessler2010a} have released a public dataset called the Supernova Photometric Classification Challenge (SNPCC) \footnote{\url{http://www.hep.anl.gov/SNchallenge/}}, which contains a mixture of simulated flux measurements with realistic observation conditions such as sky noises, point spread functions (PSFs) and atmospheric transparency.
  This dataset became a defacto-standard dataset for photometric classification, and many subsequent studies \cite{Richards2012,Karpenka2013,Lochner2016,Charnock2016} examined their own methods using this dataset.
  However, this approach requires multi-epoch observations and a photometric redshift (or simply called photo-z) to fit the observations to light curves.
  Poznanski \etal \cite{Poznanski2007} have tried single-epoch classification, however, they required precise estimations of the photometric redshifts of supernovae, and the classification performance was not as good as that of other photometric approaches.

  Our proposed method can identify SNIa samples with only single-epoch observations and no redshift by integrating the state-of-the-art computer vision methodology into the photometric approach widely used in cosmology.





\section{Building a dataset}
\label{sec:data}


  We first obtain a publicly available archive called the Cosmic Evolution Survey (COSMOS) archive \footnote{\url{http://irsa.ipac.caltech.edu/data/COSMOS/}}, which includes images, spectra and catalogs of galaxies.
  We use only the galaxy catalogs with $0.1\le$photo-z$\le 2.0$ in this archive, and prepare images of galaxies registered in the catalogs from our past observation archives.

\begin{figure}[t]
  \begin{center}
    \includegraphics[mediaboxonly,width=0.485\linewidth]{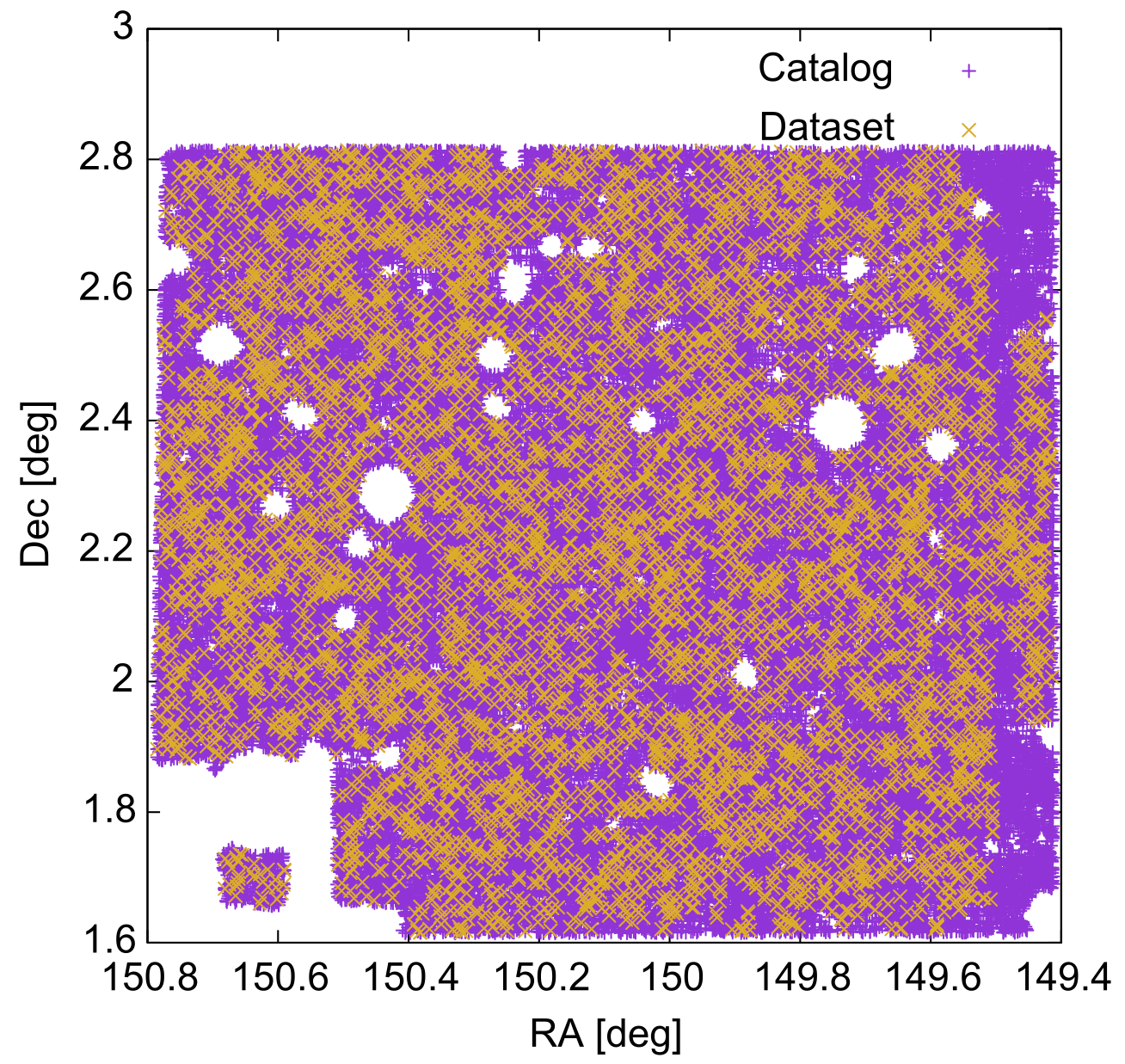}
    \includegraphics[mediaboxonly,width=0.485\linewidth]{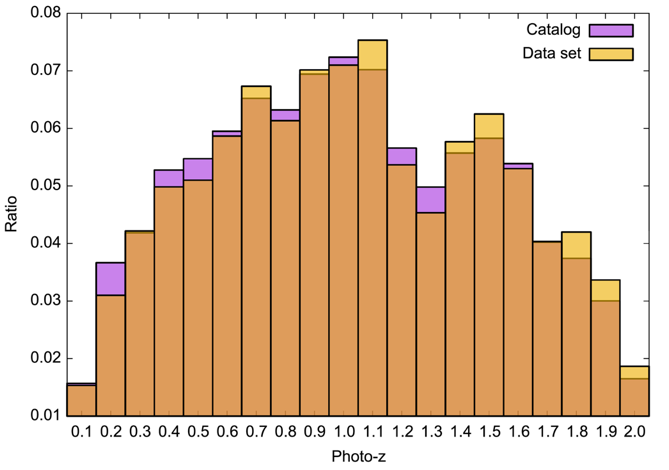}
  \end{center}
  \caption{%
    Left: Spatial distribution of host galaxies in the catalog (purple) and the dataset (orange).
    Right: Distributions of the redshifts of the galaxies in the catalog (purple) and the dataset (orange).
  }
  \label{fig:distribute_galaxy}
\end{figure}

\begin{figure}[t]
  \begin{center}
    \includegraphics[mediaboxonly,width=0.485\linewidth]{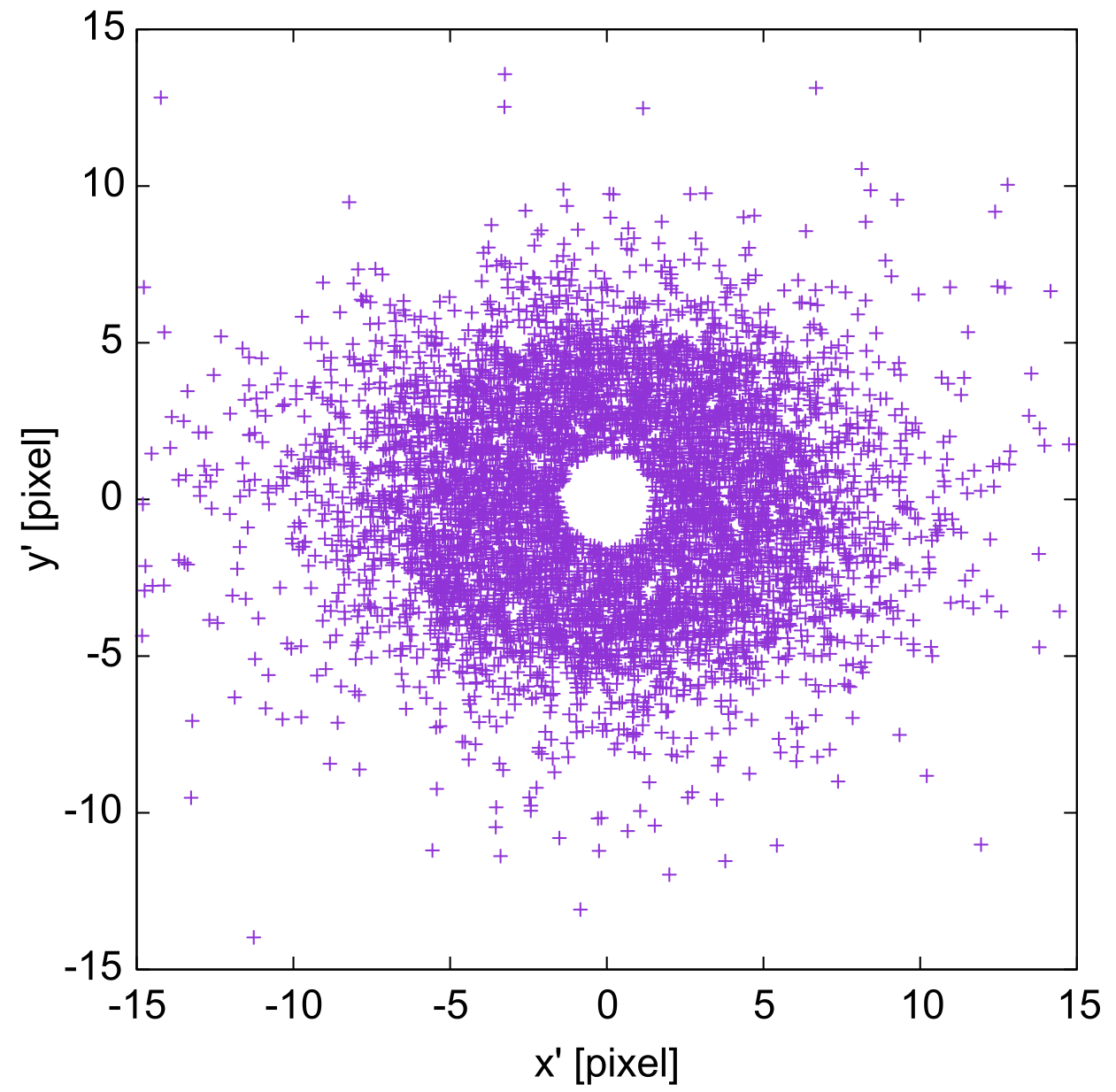}\hspace{7pt}
    \includegraphics[mediaboxonly,width=0.485\linewidth]{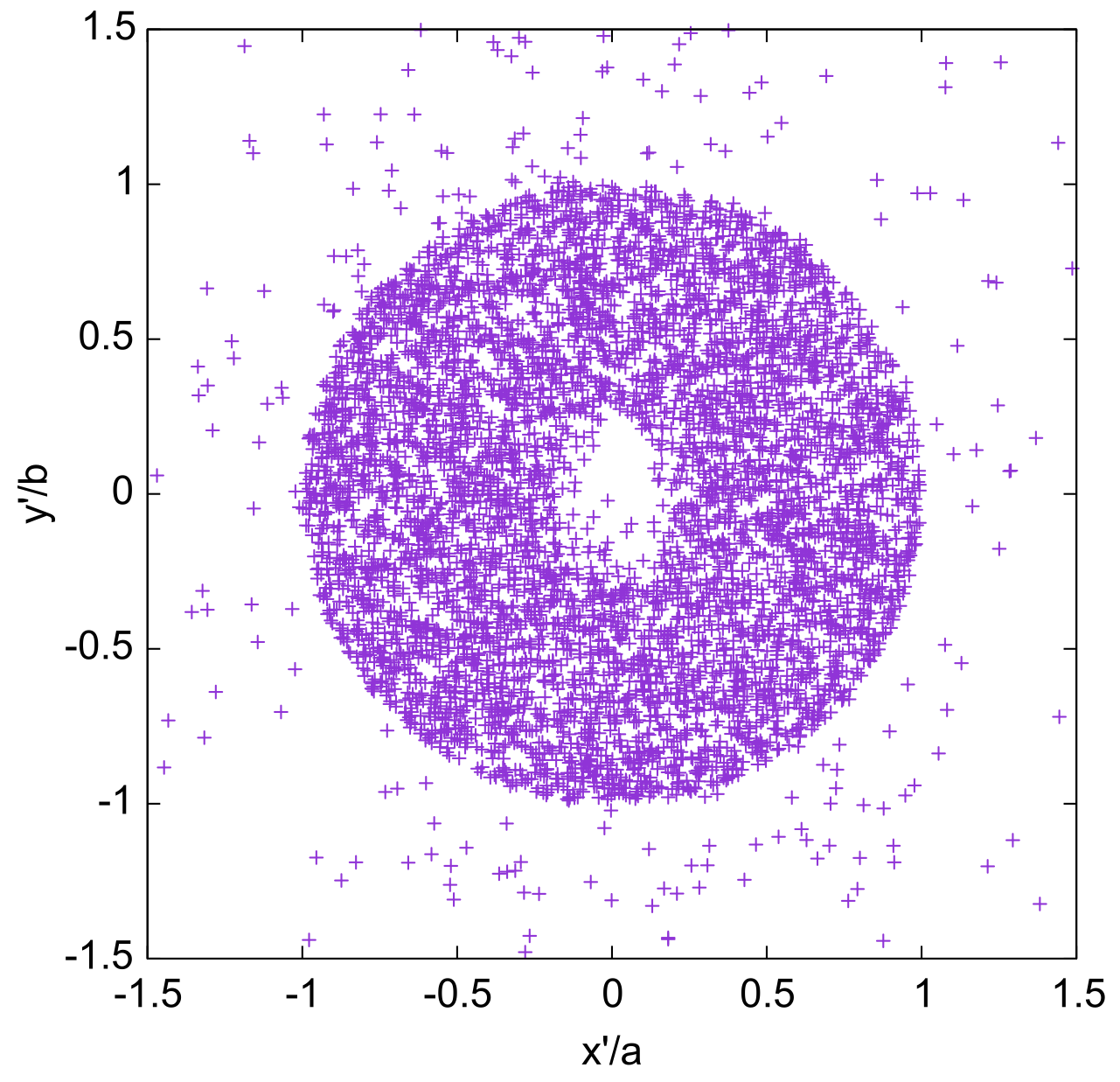}
  \end{center}
  \caption{%
    Spatial distribution of supernovae in the dataset, centered at the position of the host galaxies.
    (Left) Raw distribution, (right) normalized by the size of the host galaxies.
  }
  \label{fig:pos_supernovae}
\end{figure}

  We next select a galaxy from the catalog and determine the position of a supernova whose host galaxy is the selected one.
  The position is randomly selected from an ellipsoidal region fitted to the host galaxy as shown in Figure \ref{fig:pos_supernovae}.
  Figure \ref{fig:distribute_galaxy} left shows the spatial distribution of the host galaxies in the COSMOS area, and the right shows the distributions of the photometric redshifts of the host galaxies.
  It can be seen that galaxies in both the catalog and the dataset cover almost the entire COSMOS area of interest.


  The third step is to generate the light curve of a supernova.
  The parameters (type, stretch and color) of a supernova are randomly selected that follow the already known distributions \cite{Mosher2014}. 

  The last step before embedding the supernova in a galaxy image is to define the observation schedules and explosion date of the supernova.
  Schedules for broad-band photometric surveys are usually fixed in advance.
  Therefore, once we have determined when the supernova reaches the peak of luminance, the flux of a supernova for every band on every observation date can be systematically computed from the light curve.
  For this dataset, we arrange the observation schedule so that no more than 2 band images are taken on the same day and every band has 4 observations in total.

\begin{figure}[t]
  \begin{center}
    \includegraphics[mediaboxonly,width=0.25\linewidth]{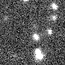}\hspace{10pt}
    \includegraphics[mediaboxonly,width=0.25\linewidth]{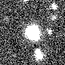}\hspace{10pt}
    \includegraphics[mediaboxonly,width=0.25\linewidth]{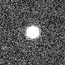}\\
    \vspace{5pt}
    \includegraphics[mediaboxonly,width=0.25\linewidth]{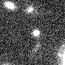}\hspace{10pt}
    \includegraphics[mediaboxonly,width=0.25\linewidth]{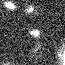}\hspace{10pt}
    \includegraphics[mediaboxonly,width=0.25\linewidth]{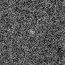}
  \end{center}
  \caption{%
    Image examples in the dataset, low photo-z $\ll 1.0$ (top) and high photo-z $>1.0$ (bottom) samples,
    a reference (left) and simulated observation (middle) image,  their difference (right).
  }
  \label{fig:data_examples}
\end{figure}

  Here, we are ready to embed the supernova in a galaxy image.
  A $65\times 65$ region is cropped from large format imaging data, and the supernova is embedded in the cropped image with an appropriate PSF, according to already determined parameters such as the position and luminance of the supernova.
  We have simulated fluctuations in observations conditions such as weathers by using the images of the same galaxy taken on different days for different observation schedules.
  Image examples can be seen in Figure \ref{fig:data_examples}.

  With this procedure, we obtain 20 observations images (5 bands $\times$ 4 observations per band, supernovae embedded), 5 reference images (no supernovae embedded) and the light curve of the embedded supernova, which we regard as a tuple of observation images, reference images and a light curve as a dataset sample.
  We have generated 6,000 SNIa and 6,000 non-SNIa (Ib, c, IIL, IIN, IIP) samples.

\section{Proposed model}
\label{sec:model}

\begin{figure}[t]
  \begin{center}
    \includegraphics[mediaboxonly,width=0.985\linewidth]{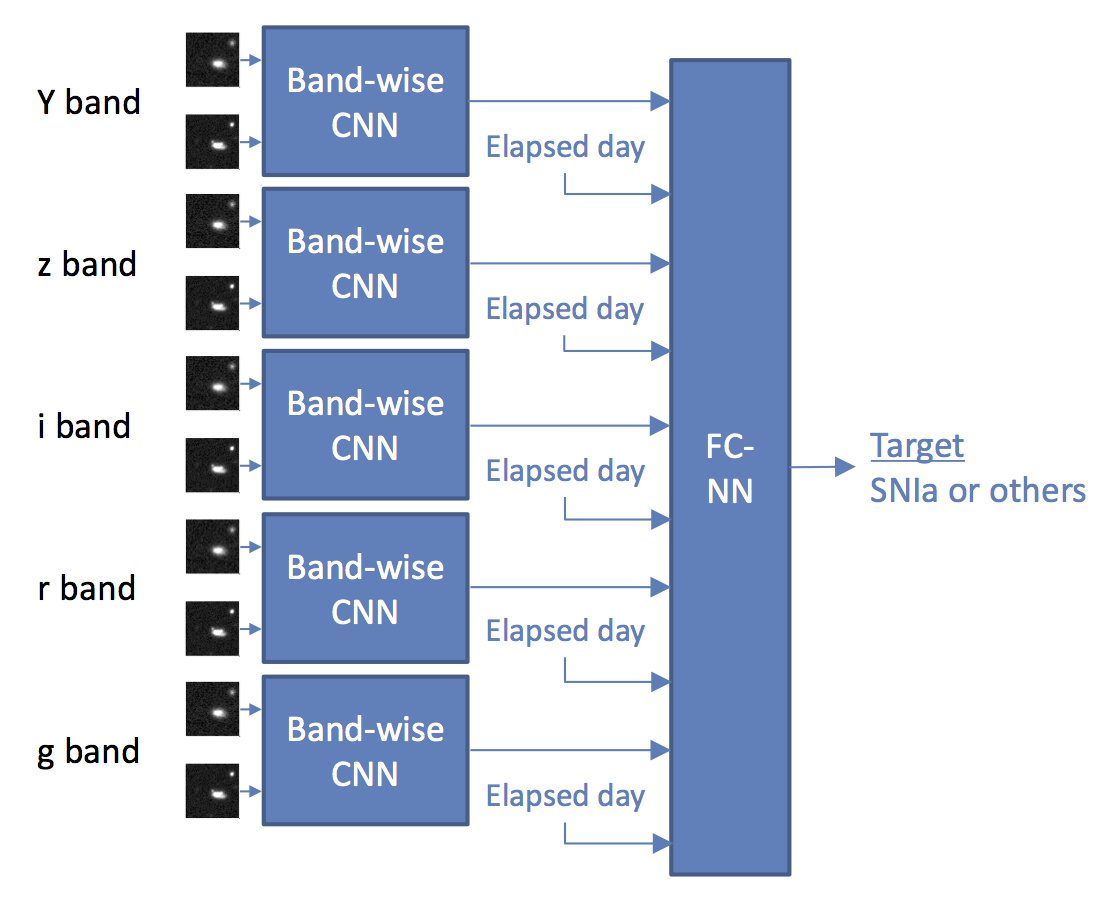}
    \end{center}
  \caption{Proposed model for supernova classification}
  \label{fig:model_unified}
\end{figure}

  Figure \ref{fig:model_unified} shows the framework of our proposed model, which consists of 5 band-wise CNNs with a pair of image observations as inputs and a fully connected NN with the estimated flux and the observation dates as inputs.

\begin{figure}[t]
  \begin{center}
    \includegraphics[mediaboxonly,width=0.985\linewidth]{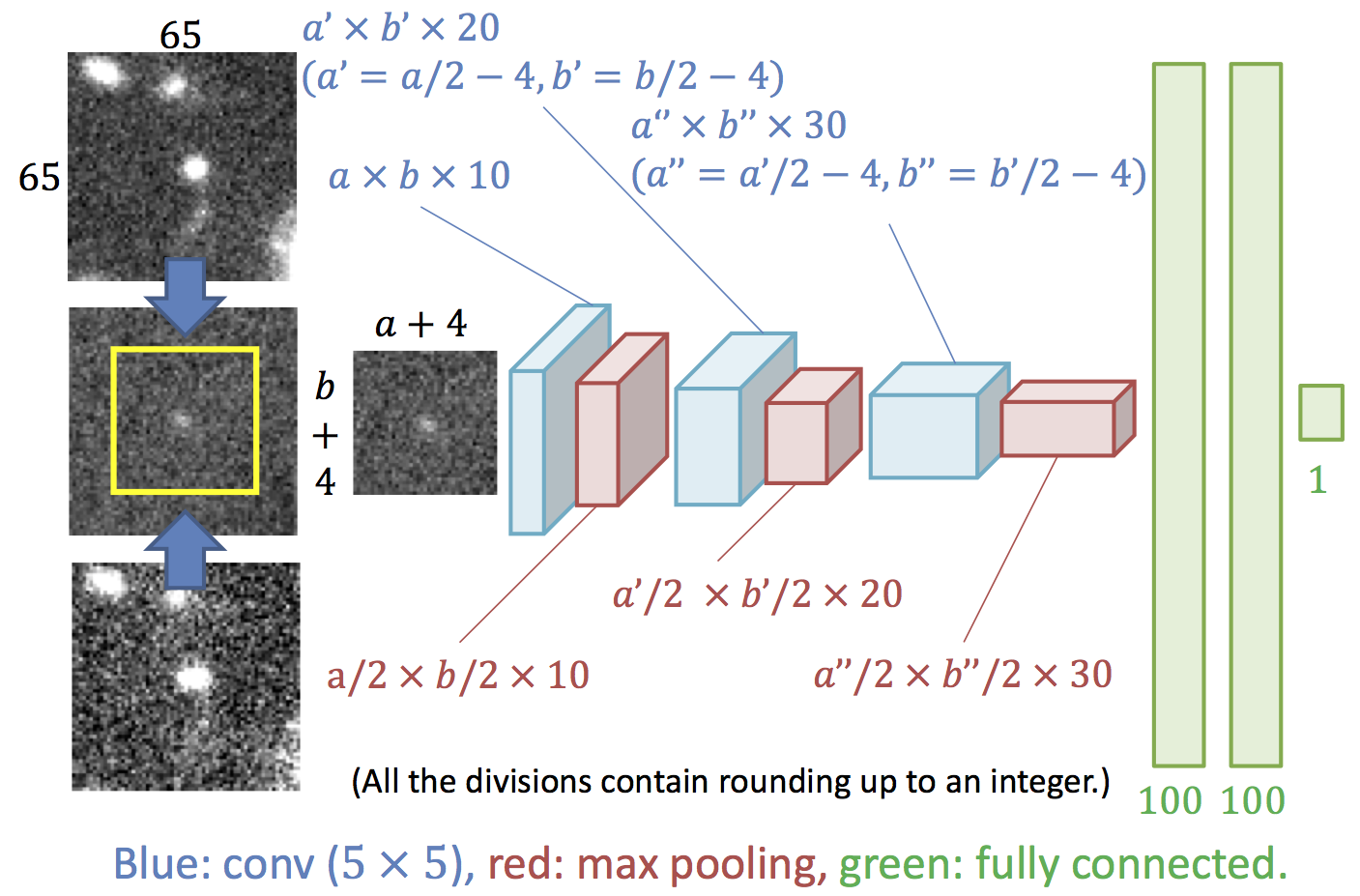}
  \end{center}
  \caption{Detailed structure of band-wise CNN for flux estimation}
  \label{fig:model_lightcurve}
\end{figure}

  The first part, a band-wise CNN, estimates the flux of the supernova candidates from a pair of reference and observation images of the same band.
  All the parameters of the band-wise CNNs are shared with all the bands.
  Figure \ref{fig:model_lightcurve} shows the detailed structure of our band-wise CNN for estimating the fluxes of supernovae.
  More specifically, this model estimates stellar magnitude (or simply magnitude) instead of flux, which can be computed as
\[
  {\rm mag} = -2.5\log_{10}{\rm flux} + 27.0.
\]
meaning that a small magnitude means a bright object.

  A difference image is first computed from an input pair of images.
  Since we want to obtain the magnitude, we employ a logarithm of the pixel values of the difference image.
  In specific, for a given pixel value $x$, we compute
\[
  y = {\rm sgn}(x)\log_{10}(x+1),
\]
where ${\rm sgn}(x)$ is a sign function that returns $1$ and $-1$ for positive and negative inputs, respectively.

  The difference image is then cropped and fed to 3-layer convolution modules, each of which consists of a $5\times 5$ convolution filter, batch normalization \cite{batchNormalization}, parametric ReLU and $2\times 2$ max pooling.
  Of those, the max pooling layer is the most important, since every observation contains no more than 1 supernova.
  The numbers of channels are set at 10, 20 and 30 for the 1st, 2nd and 3rd layers, respectively.
  After the image has passed through 3-layer fully connected modules, we can obtain a magnitude of the supernova.

  The second part, a fully connected NN, integrates the concept of photometric supernova classification with image-based classification.
  Therefore, the task of the second part is a binary (SNIa or not) classification with 10-dimensional light curve features composed of the estimated flux and the observation date for each band.
  The idea is that we do not have to fit flux observations to light curves if we aim at supernova type classification and a single-epoch observation is sufficient for this purpose.
  The network comprises the first fully-connected layer, 2 highway networks \cite{Srivastava2015} and the last fully connected layer for the output.

  For the testing stage, we need (1) a pair of reference and observation images for each band (2 images $\times$ 5 bands $=$ 10 images in total) and (2) the observation dates, and therefore flux measurements are no longer required.

\section{Experiments}
\label{sec:exp}

  We used an original synthetic dataset to train and test our models presented in Section 3. 
  It should be noted that almost all the previous researches for supernova classification utilized synthetic datasets and we also follow this strategy, since the final decision of supernova type classification relies on spectroscopic observations and due to this bottleneck only a few real-world examples can be obtained.
  As shown in Section \ref{sec:data}, the dataset contains 12,000 samples in total, 6,000 of which are SNeIa and the rest are other types of supernovae.
  In the training process, 80\% (9,600 samples) were used for training, 10\% (1,200 samples) for validation, and 10\% (1,200 samples) were kept for testing.
  To simulate single-epoch observations, we split each sample into 4 subsets, each of which included a pair of reference and observations samples for each band, the corresponding light curve data and observation date.
  For the flux estimation, we further divided each subset of a single-epoch observation into 5 pairs of images, each corresponding to a band.



{\footnotesize
\begin{table}[t]
  \centering
  \caption{Mean loss for image sizes ($10^{-3}$)}
  \label{table:size_lightcurve}
  \begin{tabular}{|l||c|c|c|} \hline
    Size          & Train loss       & Val loss          & Test loss \\ \hline \hline
    $36\times 36$ & $10.506\pm 0.493$ & $8.684\pm 1.003$ & $11.468$  \\ \hline
    $44\times 44$ & $10.269\pm 0.515$ & $8.886\pm 1.648$ & $ 8.077$  \\ \hline
    $52\times 52$ & $10.149\pm 0.500$ & $8.818\pm 1.758$ & $ 8.661$  \\ \hline
    $60\times 60$ & $ 9.763\pm 0.921$ & $7.668\pm 0.610$ & $ 7.541$  \\ \hline
    $65\times 65$ & $10.201\pm 0.508$ & $9.666\pm 2.432$ & $ 7.692$  \\ \hline
  \end{tabular}
\end{table}
}

{\footnotesize
\begin{table*}[t]
  \caption{%
    Comparisons with existing methods.
  }
  \label{table:compare_classify}
  \begin{center}
    \begin{tabular}{|l||l|l|c|} \hline
      Method & Features & Dataset & AUC \\ \hline \hline
      \multirow{3}{*}{Poznanski2007 \cite{Poznanski2007}} & Single-epoch + redshift & SNLS (partly real) & accuracy$=0.97$ \\ \cline{2-4}
       & Single-epoch, w/o redshift & SNLS (partly real) & accuracy$=0.60$ \\ \cline{2-4}
       & Single-epoch + redshift & Synthetic & accuracy$\ll 0.9$ \\ \hline
      \multirow{2}{*}{Lochner2016 \cite{Lochner2016}} & Multi-epoch (4-40) + redshift & SNPCC (synthetic) & 0.984 \\ \cline{2-4}
       & Multi-epoch (4-40), w/o redshift & SNPCC (synthetic) & 0.976 \\ \hline
      Moller2016 \cite{Moller2016} & Multi-epoch (not disclosed) + redshift & SNLS3 (synthetic) & 0.97 \\ \hline
      \multirow{2}{*}{Charnock2016 \cite{Charnock2016}} & Multi-epoch (4-40) + redshift & SNPCC (synthetic) & 0.981 \\ \cline{2-4}
       & Multi-epoch (4-40), w/o redshift & SNPCC (synthetic) & 0.981 \\ \hline
      \multirow{2}{*}{\textbf{Proposed}} & \textbf{Single-epoch, w/o redshift} & \textbf{Synthetic} & \textbf{0.958} \\ \cline{2-4}
       & \textbf{Multi-epoch (4), w/o redshift} & \textbf{Synthetic} & \textbf{0.995} \\ \hline
    \end{tabular}
  \end{center}
\end{table*}
}

\begin{figure}[t]
  \begin{center}
    \includegraphics[mediaboxonly,width=0.985\linewidth]{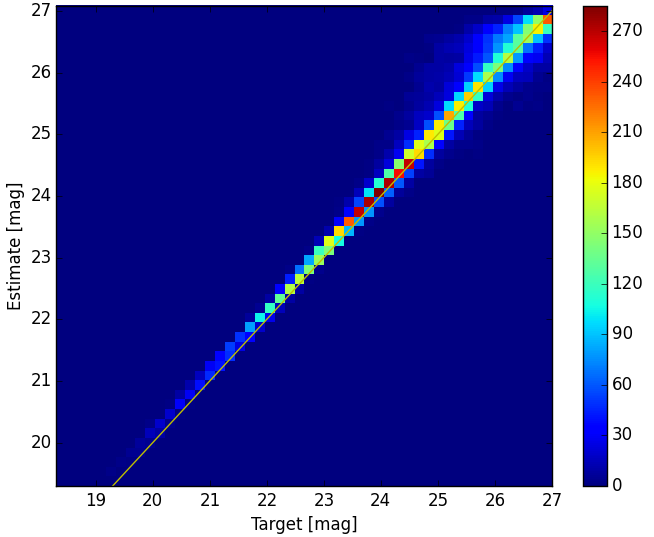}
  \end{center}
  \caption{%
    Ground-truth vs. estimated light magnitudes for test data. The yellow line shows target$=$estimated.
  }
  \label{fig:heatmap_lightcurve}
\end{figure}

\begin{figure}[t]
  \begin{center}
    \includegraphics[mediaboxonly,width=0.985\linewidth]{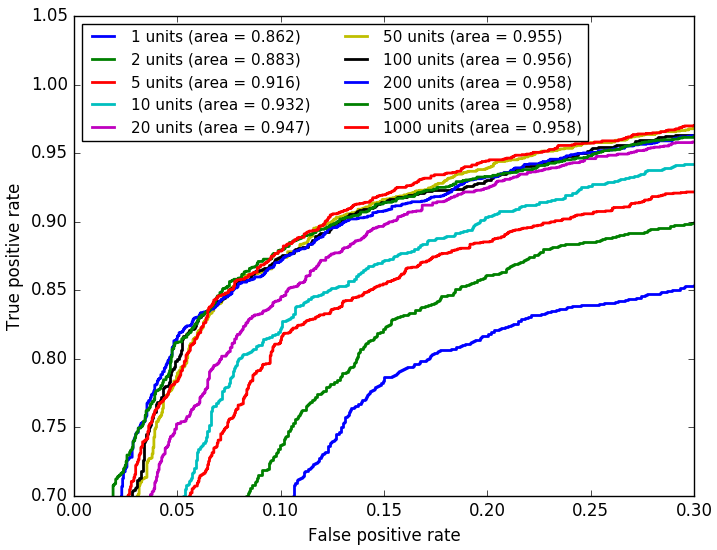}
  \end{center}
  \caption{%
    Classification performances with ground-truth light curve features for various numbers of units in the model, measured in terms of ROC curve.
  }
  \label{fig:roc_classify}
\end{figure}

\begin{figure}[t]
  \begin{center}
    \includegraphics[mediaboxonly,width=0.985\linewidth]{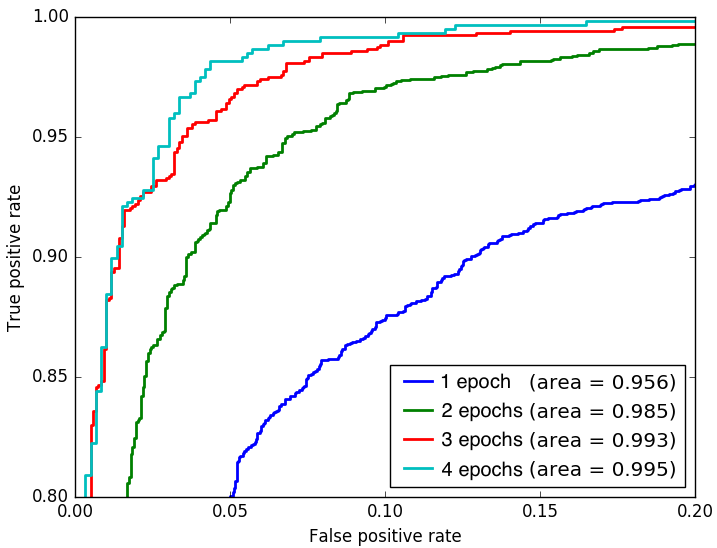}
  \end{center}
  \caption{%
    Relationships between observation epochs and classification performances, measured in terms of the ROC curve.
  }
  \label{fig:roc_classify_epoch}
\end{figure}




\begin{figure}[t]
  \begin{center}
    \includegraphics[mediaboxonly,width=0.985\linewidth]{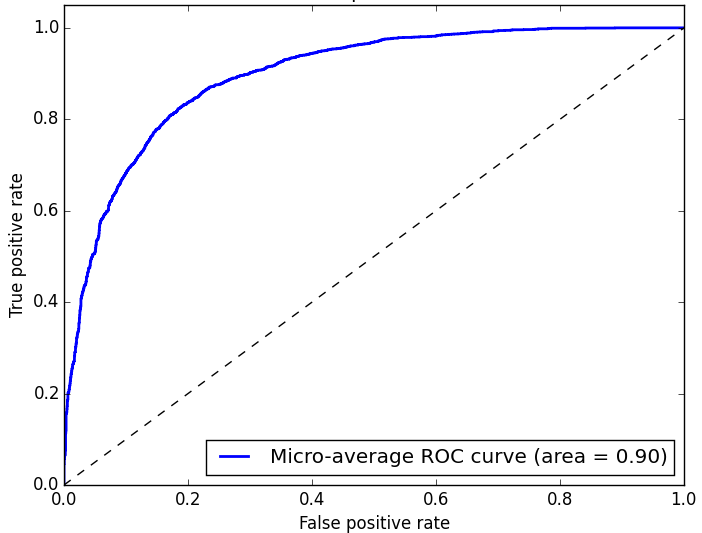}
  \end{center}
  \caption{%
    Classification performance of the joint model, measured by using the ROC.
  }
  \label{fig:roc_unified}
\end{figure}

\begin{figure}[t]
  \begin{center}
    \includegraphics[mediaboxonly,width=0.985\linewidth]{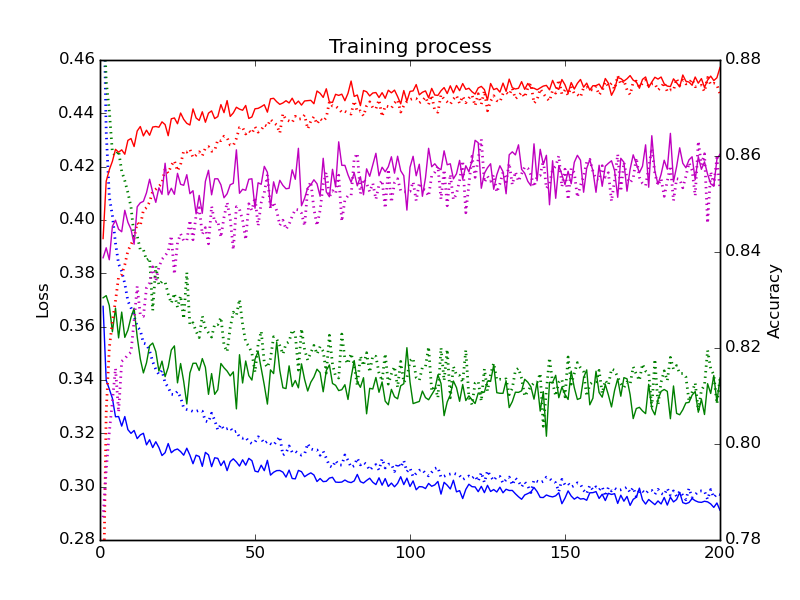}
    \end{center}
  \caption{%
    Training process for training from scratch (dashed) and our fine tuning (solid).
    (Blue) Training and (green) validation loss, (red) training and (purple) validation accuracy.
  }
  \label{fig:loss_unified}
\end{figure}




  First, we present the results for estimating the fluxes of supernova candidates with band-wise CNNs.
  Figure \ref{fig:heatmap_lightcurve} left shows the test performances measured by the mean squared loss between ground-truth and estimated light magnitudes.
  We here used the images of size $60\times 60$ as inputs. 
  This indicates that our proposed model well estimated the fluxes of supernovae about the mean estimation error of $0.087$ light magnitudes.
  This also indicates that our model produces relatively high variances for dark (i.e. large magnitudes) objects and tends to estimate the fluxes of bright objects slightly darker than the targets.

  Table \ref{table:size_lightcurve} shows the mean training, validation and test losses for several image sizes.
  We can see a trend that larger images yield better performance, and thus precise flux estimation can be achieved by considering not only central supernova regions but also background regions.


  Next, we present the results for classification with ground-truth light curve features and a fully connected NN.
  Figure \ref{fig:roc_classify} right represents the classification performance of the proposed method measured by using the ROC curve for various numbers of units in the NN. 
  As you can see, our proposed method successfully selected SNIa samples with AUC of $0.958$. 
  This also indicates that 100 units is sufficient to yield better classification performances.
  Figure \ref{fig:roc_classify_epoch} shows relationships between observation epochs and classification performance.
  This indicates that increase of observation epochs greatly improved the classification performance, however, single-epoch classification yielded sufficiently good classification performance.

  Table \ref{table:compare_classify} shows comparisons with recent methods.
  It should be noted that we used a different dataset from other existing researches, since the objective of our method is to classify supernovae directly from observation images without any complex flux measurements and existing datasets do not contain observation images.
  This table demonstrates the effectiveness of our proposed method against existing photometric-based methods.
  More specifically, (1) under the same conditions of features (single-epoch observations and no redshifts), our method greatly outperformed the method by Poznanski \etal \cite{Poznanski2007}, (2) our method was comparable to those with multi-epoch observations even if we had only single-epoch features, and (3) when multi-epoch features are available, our method outperformed all the existing methods.




  Finally, we present the results for a joint model that combines band-wise DCNNs and fully connected layers integrating all the CNN outputs.
  Figure \ref{fig:roc_unified} represents the classification performance measured by the ROC curve. 
  As you can see, our joint model achieved the AUC of $0.897$.

  We have compared our fine-tuning strategy with training the joint model from scratch.
  Figure \ref{fig:loss_unified} presents the result, which indicates that our fine-tuning strategy outperformed training from the scratch and converged much faster.



\section{Conclusion}
\label{sec:conclude}

  In this paper, we proposed a novel method for classifying SNeIa simply from single-epoch observation images by effectively integrating the state-of-the-art computer vision methodology into the photometric approach widely used in cosmology.
  With our proposed method, a band-wise CNN estimates the luminance of supernovae from telescope images, and another fully connected NN classified the supernovae with the estimated luminances and observation dates as features.
  Experimental results demonstrated the effectiveness of the proposed method by comparison with existing photometric classification methods with many observations.

  The dataset we have built for this work is the largest of those containing image and light curves of supernovae and will contribute to further improvements in supernova detection and classification.

\ifCLASSOPTIONcompsoc
  \section*{Acknowledgments}
\else
  \section*{Acknowledgment}
\fi

  This work is supported by Core Research for Evolutionary Science and TEchnology (CREST), Japan Science and Technology (JST).

\bibliographystyle{ieee}
\bibliography{supernova}

\end{document}